\documentclass[a4paper,11pt]{article}
\usepackage{pos}
\pdfoutput=1

\title{Parton Distribution Functions in the Schwinger Model with Tensor Networks}

\author[a,b]{Mari Carmen Bañuls}
\author[c]{Krzysztof Cichy}
\author[d,e]{C.-J. David Lin}
\author*[d]{Manuel Schneider}

\affiliation[a]{Max-Planck-Institut f{\"u}r Quantenoptik,\\ Hans-Kopfermann-Str. 1, 85748 Garching, Germany}
\affiliation[b]{Munich Center for Quantum Science and Technology (MCQST),\\ Schellingstr. 4, 80799 Munich, Germany}

\affiliation[c]{Faculty of Physics and Astronomy, Adam Mickiewicz University,\\ ul.\ Uniwersytetu Poznańskiego 2, 61-614 Poznań, Poland}

\affiliation[d]{Institute of Physics, National Yang Ming Chiao Tung University,\\
	1001 University Road, Hsinchu 30010, Taiwan}

\affiliation[e]{Centre for High Energy Physics, Chung-Yuan Christian University,\\
        200 Chung-Pei Road, Chung-Li District, Taoyuan 320314, Taiwan}

\emailAdd{manuel.schneider@nycu.edu.tw}

\abstract{Parton distribution functions (PDFs) describe universal properties of bound states and allow us to calculate scattering amplitudes in processes with large momentum transfer. Calculating PDFs involves the evaluation of matrix elements with a Wilson line in a light-cone direction. In contrast to Monte Carlo methods in Euclidean spacetime, these matrix elements can be directly calculated in Minkowski-space using the Hamiltonian formalism. The necessary spatial- and time-evolution can be efficiently applied using established tensor network methods. We present PDFs in the Schwinger model calculated with matrix product states.}

\FullConference{%
  The 41st International Symposium on Lattice Field Theory, LATTICE2024\\
  28th July - 3rd August 2024\\
  University of Liverpool, United Kingdom
}

\renewcommand{\hookAfterAbstract}{%
	\par\bigskip
	\textsc{ArXiv ePrint}:
	\href{https://arxiv.org/abs/2409.16996}{2409.16996}
}

\hypersetup{%
		pdfkeywords        = {PDF} {MPS} {lightcone} {Schwinger},
		pdfsubject         = {Parton distribution functions (PDFs) describe universal properties of bound states and allow us to calculate scattering amplitudes in processes with large momentum transfer. Calculating PDFs involves the evaluation of matrix elements with a Wilson line in a light-cone direction. In contrast to Monte Carlo methods in Euclidean spacetime, these matrix elements can be directly calculated in Minkowski-space using the Hamiltonian formalism. The necessary spatial- and time-evolution can be efficiently applied using established tensor network methods. We present PDFs in the Schwinger model calculated with matrix product states.},
	pdftitle           = {Parton Distribution Functions in the Schwinger Model with Tensor Networks}
	pdfcreator         = {Mari Carmen Bañuls, Krzysztof Cichy, C.-J. David Lin, and Manuel Schneider},
	pdfauthor				   = {Mari Carmen Bañuls, Krzysztof Cichy, C.-J. David Lin, and Manuel Schneider},
	pdflang            = {en},
	bookmarksopen      = true,
	bookmarksopenlevel = 3,
	hypertexnames      = false,
	linktocpage        = true,
	plainpages         = false,
	breaklinks
}

\usepackage[english]{babel}
\usepackage[utf8]{inputenc}
\usepackage[T1]{fontenc}
\usepackage[babel,final]{microtype}
\usepackage{hyperref}			
\usepackage{url}
\usepackage{hyphsubst}		
\usepackage{float}				

\usepackage{amsfonts}
\usepackage{amsmath}
\usepackage{MnSymbol}	
\usepackage{slashed}	
\usepackage{braket}

\usepackage{xspace}		
\usepackage{tikz}

\usepackage[numbers,sort&compress]{natbib}
\usepackage{cleveref}
\crefname{section}{sec.}{sections}
\Crefname{section}{Section}{Sections}
\crefname{appendix}{app.}{appendices}
\Crefname{appendix}{Appendix}{Appendices}
\crefname{figure}{fig.}{figs.}
\Crefname{figure}{Figure}{Figures}
\crefname{equation}{eq.}{eqs.}
\Crefname{equation}{Equation}{Equations}

\usepackage[width=.9\textwidth,font=small,hypcap=true]{caption}
\usepackage[hypcap=true]{subcaption}
\newcommand{\cor}{\ensuremath{\mathcal{M}}\xspace}
\newcommand{\te}[1][]{\ensuremath{e^{-i H {#1}}}\xspace}
\newcommand{\tedag}[1][]{\ensuremath{e^{i H {#1}}}\xspace}
\newcommand{\z}{\ensuremath{d}\xspace}
\newcommand{\zphys}{\ensuremath{z}\xspace}
\newcommand{\bjorx}{{\ensuremath{\xi}}\xspace}
\newcommand{\pdf}{\ensuremath{f_{\psi}\left(\bjorx\right)}\xspace}
\newcommand{\antipdf}{\ensuremath{f_{\overline{\psi}}\left(\bjorx\right)}\xspace}

\newcommand*{\doi}[1]{
	\IfSubStr{#1}{ and }{\StrCut{#1}{ and }{\csA}{\csB}dois: \href{https://dx.doi.org/\csA}{\csA} and \href{https://dx.doi.org/\csB}{\csB}}
		{\href{https://dx.doi.org/#1}{doi: #1}}
}

\renewcommand{\Re}{\operatorname{Re}}
\renewcommand{\Im}{\operatorname{Im}}

\usepackage[skins]{tcolorbox}
\usepackage{pgfplots} 

\usetikzlibrary{external}

\usetikzlibrary{arrows,backgrounds,fit,calc,patterns,shapes}

\usetikzlibrary{plotmarks}
\pgfplotsset{compat=newest}
\usetikzlibrary{plotmarks}
\usetikzlibrary{arrows.meta}
\usepgfplotslibrary{patchplots}

\tikzset{every node/.style={sloped,allow upside down},baseline={([yshift=-0.5ex]current bounding box.center)},inner sep=.3mm,x=1cm,y=1cm}

\definecolor{mycolor1}{rgb}{0.400,0.612,0.086}%
\definecolor{mycolor2}{rgb}{0.353,0.137,0.549}%
\definecolor{tensorblue}{rgb}{0.8,0.8,1}
\tikzset{tengreen/.style={fill=green!50!black!50}}

\tikzset{ket/.style={circle,draw=black,thick,fill=tensorblue}}
\tikzset{timeline/.style={->,line width=0.5mm,color=mycolor1}}
\tikzset{spaceline/.style={->,line width=0.5mm,color=mycolor2}}

\begin{document}
\maketitle

\newpage

\section{Introduction}

One of the central aims of particle physics is to achieve detailed understanding of the inner structure of hadrons, such as the nucleon or the pion. This structure is governed by the strong interaction and described by quantum chromodynamics (QCD). Due to the non-Abelian nature of QCD and the running of the strong coupling, the structure of hadrons is very intricate and can be quantified in terms of different kinds of partonic functions. The simplest of these are parton distribution functions (PDFs), which represent the fraction of momentum carried by the hadron's constituents, quarks and gluons~\cite{collins_2011}.  PDFs are given in terms of light-front matrix elements and thus, they cannot be directly accessed in Euclidean lattice QCD. Instead, they are being computed indirectly, using factorization-based approaches. The most widely used of these are  quasi-distributions \cite{Ji:2013dva,Ji:2014gla} and pseudo-distributions \cite{Radyushkin:2017cyf}. However, other approaches also exist, such as the heavy-quark operator product expansion \cite{Detmold:2005gg, Detmold:2021uru}, current-current correlators \cite{Braun:2007wv,Ma:2017pxb}, the hadronic tensor \cite{Liu:1993cv} and the Compton amplitude \cite{Chambers:2017dov}. For a review about the recent progress, see Refs.~\cite{Cichy:2018mum,Radyushkin:2019mye,Ji:2020ect,Constantinou:2020pek,Cichy:2021lih,Cichy:2021ewm}.

However, it would be desirable to extract partonic functions more directly from Minkowski-space matrix elements. This can in principle be achieved using tensor network state (TNS) techniques in the Hamiltonian framework.
TNSs are efficient ansatzes for the wave function, which can be subjected to real-time evolution, thus giving access to light-front matrix elements. TNS methods are particularly successful in low dimensions, where the approach of matrix product states (MPS) \cite{verstraete04dmrg,perez07mps,MPS} is applicable and their suitability for equilibrium states of lattice gauge theories (LGTs) has been systematically established (see e.g.~\cite{Banuls2020ropp} and references therein).

Here we present an exploratory study of this possibility, focused on the Schwinger model \cite{schwinger62}, i.e.\ quantum electrodynamics (QED) in 1+1 dimensions. The model shares several non-trivial properties with QCD, such as confinement and asymptotic freedom. Originally, it was introduced as a model of non-perturbative generation of a mass gap. The lightest composite state of this model, the vector meson, can be thought of as an analogue of a hadron in QCD. The vector meson is a bound state of fermions and antifermions, which can be considered as partons that are described by PDFs~\cite{SchwingerPartonPerturbative}.

Whereas earlier calculations of Schwinger model PDFs in the Hamiltonian formalism exist, they were limited to exact diagonalization in a small momentum lattice~\cite{Schwinger_infiniteMomentum,SchwingerPDF_fastMovingFrame,SchwingerPDF_fastMovingFrameProceedings} or numerical integrations in the approximation of small fermion numbers~\cite{SchwingerMassivePDF}. In contrast, the TNS approach, which we apply to PDFs for the first time, allows treating very large systems, and a systematic improvement for convergence.

\section{The Schwinger Model}
The massive Schwinger model is described by the Lagrangian
\begin{equation}
	{\cal{L}} = \overline{\psi} (i \slashed{\partial} - g \slashed{A} - m )\psi - \frac{1}{4} F_{\mu\nu} F^{\mu \nu},
\end{equation}
where $\psi$ are Dirac fermions coupled to the gauge field $A$ with a coupling constant $g$, and $m$ is the fermion mass. The electromagnetic tensor is defined as $F_{\mu\nu} = \partial_\mu A_\nu - \partial_\nu A_\mu$.

The Hamiltonian can be obtained by a Legendre transformation~\cite{Banks1976,Hamer1997,Schwinger_massSpectrum,Schwinger_QC}. We work in the temporal gauge, $A^0=0$. The model is evaluated on a lattice with lattice spacing $a$ and with open boundary conditions. We use Kogut-Susskind staggered fermions, where the upper (lower) components of the Dirac fermions are placed on even (odd) sites of the lattice~\cite{staggeredFermions}. A Jordan-Wigner transformation maps the fermionic degrees of freedom to  spin variables. The Hamiltonian for an $N$-site system becomes \footnote{We choose $\gamma^0 = \sigma^z$, $\gamma^1 = i \sigma^y$ with Pauli matrices $\sigma^x$, $\sigma^y$ and $\sigma^z$; $\sigma^{\pm}=\frac{1}{2}\left(\sigma^{x}\pm i\sigma^{y}\right)$; $\gamma^\pm = \frac{1}{\sqrt{2}} \left( \gamma^0 \pm \gamma^1 \right)$.}
\begin{equation}
	H_\text{spin}
	=\frac{1}{2 a}\sum_{n=0}^{N-2} \left (  \sigma^{+}_n \sigma^{-}_{n+1} + \sigma^{-}_{n+1} \sigma^{+}_{n}\right ) 
	+\frac{m_\text{lat}}{2}\sum_{n=0}^{N-1} \left [ 1 + (-1)^n \sigma_n^z \right ]
	+\frac{a g^2}{2} \sum_{n=0}^{N-1} L_n^2.
\end{equation}
The electric field is described in a ladder space and the electric field operator $L_n$ counts the electric flux on a link between sites $n$ and $n+1$~\cite{Banks1976}. The additively-renormalized lattice mass $m_\text{lat} = m - \frac{ag^2}{8}$ ensures fast convergence to the continuum limit~\cite{Schwinger_latticeMass,MassRenormalizationNumerical,MassRenormalizationNumerical_PoS}.

Physical states in the Hamiltonian approach need to fulfill the Gauss law, which in spin language reads
\begin{equation}
	L_n-L_{n-1}=\frac{1}{2}\left[(-1)^n + \sigma^z_n\right] + q_n.
	\label{eq:GaussLaw}
\end{equation}
Notice that we introduced here a static charge $q_n$ at site $n$, which was not part of the original model. This will help us in \cref{sec:PDF} to calculate matrix elements with a Wilson line along the light-cone. The static charge does not have a kinetic term and enters the model only by changing the charge locally.

The electric and fermionic fields are not independent degrees of freedom because of the Gauss law. In 1+1 dimensions and with open boundary conditions, this can be used to eliminate the electric field from the Hamiltonian, which (after rescaling by a factor $\frac{2}{ag^2}$) becomes
\begin{equation}
	H = x\sum_{n=0}^{N-2} \left [ \sigma_n^+\sigma_{n+1}^- +  \sigma_n^-\sigma_{n+1}^+ \right ]+\frac{\mu}{2}\sum_{n=0}^{N-1} \left [ 1 + (-1)^n \sigma_n^z \right ] +\sum_{n=0}^{N-2} \left [ \ell +\frac{1}{2}\sum_{k=0}^n \left((-1)^k+\sigma_k^{z} + 2 q_k\right)\right ]^2.
	\label{eq:H}
\end{equation}
We defined $x = \frac{1}{a^2 g^2}$ and $\mu = \frac{2m_\text{lat}}{a g^2}$. The electric field at the left boundary $l$ is set to zero in all our calculations. We furthermore restrict ourselves to states with zero total charge.

The ground state of the Schwinger model is calculated using a variational optimization of the MPS~\cite{MPS,Schwinger_massSpectrum}. Subsequently, the ground state is projected out by a penalty term, and a second penalty term ensures that the next state belongs to the zero charge sector. In this way, we can find the first excited state as well by a variational minimization of the energy~\cite{Schwinger_massSpectrum}.  

\section{Parton Distribution Functions\label{sec:PDF}}
The PDF of a fermion $\psi$ can be calculated by a Fourier transform of light-cone matrix elements~\cite{collins_2011}:
\begin{equation}
	\pdf = \int_{-\infty}^{\infty} \frac{d\lambda}{4\pi} e^{-i \lambda \bjorx n \cdot P} \Braket{P| \overline{\psi}(\lambda n) W(\lambda n \leftarrow \vec{0}) n \cdot \gamma \psi(\vec{0})|P}_c.
	\label{eq:PDFframeIndependent}
\end{equation}
It depends only on Bjorken-$x$ (in this work denoted as $\bjorx$). $W$ is a Wilson line along the light cone.
The light-cone vector is taken to be $n^\mu = \frac{1}{\sqrt{2}} (1,1)$ in this work. The PDF can be obtained from any momentum eigenstates $\Ket{P}$ of the fermion $\psi$. Even though momentum is not well-defined for open boundary conditions, the first excited state becomes the vector boson at rest in the thermodynamic and continuum limits. We can, thus, calculate PDFs from lattices with open boundary conditions for large enough volumes and fine enough lattices. The subscript $c$ on the expectation value in \cref{eq:PDFframeIndependent} indicates that vacuum expectation values should be subtracted from matrix elements~\cite{collins_2011}.

Each Dirac fermion at a spatial position $\zphys=\z \frac{a}{2}$ relates to two sites $\z$ and $\z+1$ on the lattice with staggered fermions (\z even). Therefore, the full matrix element \cor can be calculated from four contributions on the lattice:
\begin{align}
	\cor\left(\zphys,0\right) =& \sqrt{2} \Braket{h| \bar{\psi}\left(\zphys,\zphys\right) W\left(\left(\zphys,\zphys\right) \leftarrow (0,0)\right) \gamma^- \psi(0,0)|h} \\
	=& \cor_\text{(e,e)}(\z,0) - \cor_\text{(e,o)}(\z,1) - \cor_\text{(o,e)}(\z+1,0) + \cor_\text{(o,o)}(\z+1,1).
    \label{eq:correlatorFull}
\end{align}
The bound state in the lattice theory is denoted as $\Ket{h}$. The matrix element $\cor_\text{(e,e)}$ connects site 0, which is defined as the even site in the middle of the spin chain, to a site $\z$ along the light cone. $\cor_\text{(o,o)}$ connects site $1$, which corresponds to the lower component of the Dirac fermion at the origin, to site $\z+1$. The remaining two matrix elements connect one even and one odd site. The sign convention follows from $\gamma^0 n \cdot \gamma = \gamma^0 \gamma^- = \frac{1}{\sqrt{2}} \begin{pmatrix}
    \phantom{+}1 & -1\\
    -1 & \phantom{+}1
\end{pmatrix}$. The upper and lower parts of the fermionic operator $\psi$ each have the form $\prod_{k<n}\left(-i\sigma_{k}^{z}\right) \sigma_{n}^{-}$ in the spin language, a Jordan-Wigner string and an annihilation operator.
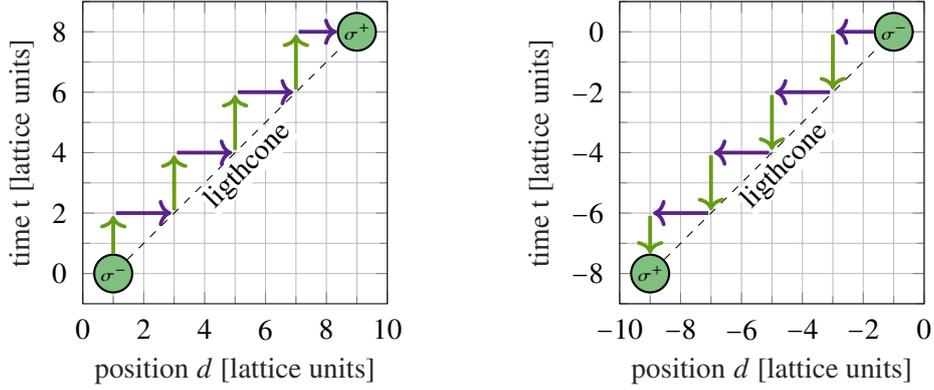
\begin{figure}[h]
	\centering
	\def\figWidth{.265/0.45*\textwidth}
	\begin{minipage}[t]{0.45\textwidth}
		\centering
\begin{tikzpicture}
	\begin{axis}[
		width=\figWidth,
		height=\figWidth,
		scale only axis,
		scale only axis,
		xmin=0,
		xmax=2*4+2,
		xlabel style={font=\color{white!15!black},yshift=-1ex},
		xlabel={position $d$ [lattice units]},
		ymin=-1,
		ymax=2*4+2-1,
		ylabel style={font=\color{white!15!black},xshift=1ex},
		ylabel={time t [lattice units]},
		axis background/.style={fill=white},
		xmajorgrids,
		ymajorgrids,
		legend style={at={(1.03,1)}, anchor=south west, legend cell align=left, align=left, draw=white!15!black},
		ylabel style={{rotate=-90}},
		scaled ticks=false,
		tick label style={/pgf/number format/fixed},
		xtick={0,2,...,10},
		ytick={0,2,...,10},
        minor tick num=1,
        xminorgrids,
		yminorgrids,
		xticklabel style={yshift=-1ex},
		yticklabel style={xshift=-1ex}
		]
		\node[ket,tengreen] (site0) at (1,0) {\tiny $\sigma^-$};
		\node[] (time0) at (0*2+1,1*2) {};
		\node[] (site1) at (1*2+1,1*2) {};
		\node[] (time1) at (1*2+1,2*2) {};
		\node[] (site2) at (2*2+1,2*2) {};
		\node[] (time2) at (2*2+1,3*2) {};
		\node[] (site3) at (3*2+1,3*2) {};
		\node[] (time3) at (3*2+1,4*2) {};
		\node[ket,tengreen] (site4) at (4*2+1,4*2) {\tiny $\sigma^+$};
		\draw[black,dashed] (site0) -- (site4) node[midway,below=1pt,opacity=1,fill=white] {ligthcone};
        \draw[timeline] (site0) -- (time0);
		\draw[spaceline] (time0) -- (site1);
        \draw[timeline] (site1) -- (time1);
		\draw[spaceline] (time1) -- (site2);
        \draw[timeline] (site2) -- (time2);
		\draw[spaceline] (time2) -- (site3);
		\draw[timeline] (site3) -- (time3);
		\draw[spaceline] (time3) -- (site4);
	\end{axis}
\end{tikzpicture}%
		\subcaption{Evolution in positive time- and spatial direction}
		\label{fig:stepwise_positive}
	\end{minipage}
	\begin{minipage}[t]{0.45\textwidth}
		\centering
\begin{tikzpicture}
	\begin{axis}[
		width=\figWidth,
		height=\figWidth,
		scale only axis,
		scale only axis,
		xmin=(-2*4-2),
		xmax=0,
		xlabel style={font=\color{white!15!black},yshift=-1ex},
		xlabel={position $d$ [lattice units]},
		ymin=-(2*4+2-1),
		ymax=1,
		ylabel style={font=\color{white!15!black},xshift=1ex},
		ylabel={time t [lattice units]},
		axis background/.style={fill=white},
		xmajorgrids,
		ymajorgrids,
		legend style={at={(1.03,1)}, anchor=south west, legend cell align=left, align=left, draw=white!15!black},
		ylabel style={{rotate=-90}},
		scaled ticks=false,
		tick label style={/pgf/number format/fixed},
		xtick={-10,-8,...,2},
		ytick={-10,-8,...,2},
        minor tick num=1,
        xminorgrids,
		yminorgrids,
		xticklabel style={yshift=-1ex},
		yticklabel style={xshift=-1ex}
		]
		\node[ket,tengreen] (site0) at (-1,0) {\tiny $\sigma^-$};
		\node[] (site1) at (-1*2-1,-0*2) {};
		\node[] (time1) at (-1*2-1,-1*2) {};
		\node[] (site2) at (-2*2-1,-1*2) {};
		\node[] (time2) at (-2*2-1,-2*2) {};
		\node[] (site3) at (-3*2-1,-2*2) {};
		\node[] (time3) at (-3*2-1,-3*2) {};
		\node[] (site4) at (-4*2-1,-3*2) {};
		\node[ket,tengreen] (site5) at (-4*2-1,-4*2) {\tiny $\sigma^+$};
		\draw[black,dashed] (site5) -- (site0) node[midway,below=1pt,opacity=1,fill=white] {ligthcone};
        \draw[spaceline] (site0) -- (site1);
		\draw[timeline] (site1) -- (time1);
		\draw[spaceline] (time1) -- (site2);
		\draw[timeline] (site2) -- (time2);
		\draw[spaceline] (time2) -- (site3);
		\draw[timeline] (site3) -- (time3);
		\draw[spaceline] (time3) -- (site4);
		\draw[timeline] (site4) -- (site5);
	\end{axis}
\end{tikzpicture}%
		\subcaption{Evolution in negative time- and spatial direction}
		\label{fig:stepwise_negative}
	\end{minipage}
	\caption{Sketch of the calculation of a matrix element in light-cone direction. The Wilson line is evolved between the coordinates of an initial $\sigma^-$ operator and those of the final $\sigma^+$ operator. On the lattice, the light-cone direction is replaced by subsequent evolutions in timelike (green vertical arrows) and spatial (violet horizontal arrows) directions.}
    \label{fig:stepwise}
\end{figure}

The Wilson line is path-dependent~\cite{WilsonLines} and needs to be evolved along a light-cone direction. This is not directly possible with a spatial lattice. We therefore replace the Wilson line by a step-wise evolution as sketched in \cref{fig:stepwise_positive} for the $\cor_\text{(o,o)}$ matrix element. After applying the initial operators at the initial site, the state is evolved for a time that corresponds to two lattice spacings $\delta t = 2a$ (with the speed of light $c=1$). Next, we evolve the electric field by two lattice sites in the spatial direction. This is followed by a time evolution again and repeated recursively until we reach the final site. The parallel transporter in time-direction becomes a time evolution of the state with the correspondingly evolved electric field~\cite{QuantumSimulation}. For the calculation of matrix elements in the negative direction, $\z < 0$, we start with a spatial evolution in the negative direction and then evolve by a negative time $\delta t = -2a$ (see \cref{fig:stepwise_negative}). Our procedure approximates the light-cone in the continuum limit.

We are working in a formulation with spin-degrees of freedom only, the gauge fields are integrated out. Therefore, we cannot evolve the electric field directly by applying operators to the state. Instead, we introduce static charges that change Gauss law locally, or, equivalently, contribute to the electric field by their charge. Whenever we apply a $\sigma^+$ or $\sigma^-$ operator, we insert a static charge at the same position with the corresponding charge, such that it cancels the effect of the $\sigma^\pm$ operator on the electric fields. The fermion number can, thus, be changed without modifying the electric flux. Then, we move the location of the static charge, which has the same effect as the application of a parallel transporter in spatial direction: changing the electric flux between two sites by one unit. Note that the static charges only enter the matrix elements by modifying the electric part of the Hamiltonian during the time evolution. Our approach shares similarities with the ancillary fermion-antifermion pairs in~\cite{QuantumSimulation}.

With these techniques, the matrix elements in positive time and spatial direction become
\begin{align}
	\cor_\text{(e,e)}(\z,0)
    =&	\Bra{h} \tedag[ t_\z] \prod_{k<\z}\left(i\sigma_{k}^{z}\right) \sigma_{\z}^{+} \te[_{\z-1} \delta t] \dots \te[_{3} \delta t]  \te[_{1} \delta t] \prod_{k'<0}\left(-i\sigma_{k'}^{z}\right) \sigma_{0}^{-} \Ket{h}, \nonumber\\
    \cor_\text{(e,o)}(\z,1)
    =&	\Bra{h} \tedag[ t_\z] \prod_{k<\z}\left(i\sigma_{k}^{z}\right) \sigma_{\z}^{+} \te[_{\z-1} \delta t] \dots \te[_{3} \delta t] \te[_{1} \delta t] \prod_{k'<1}\left(-i\sigma_{k'}^{z}\right) \sigma_{1}^{-} \Ket{h}, \label{eq:corrs}\\
    \cor_\text{(o,e)}(\z+1,0)
    =&	\Bra{h} \tedag[ t_\z] \prod_{k<\z+1}\left(i\sigma_{k}^{z}\right) \sigma_{\z+1}^{+} \te[_{\z-1} \delta t] \!\dots \te[_{3} \delta t] \te[_{1} \delta t] \prod_{k'<0}\left(-i\sigma_{k'}^{z}\right) \sigma_{0}^{-} \Ket{h}, \nonumber\\
    \cor_\text{(o,o)}(\z+1,1)
    =&	\Bra{h} \tedag[ t_\z] \prod_{k<\z+1}\left(i\sigma_{k}^{z}\right) \sigma_{\z+1}^{+} \te[_{\z-1} \delta t] \!\dots \te[_{3} \delta t] \te[_{1} \delta t] \prod_{k'<1}\left(-i\sigma_{k'}^{z}\right) \sigma_{1}^{-} \Ket{h}. \nonumber
\end{align}
Here, $\te[_{n} \delta t]$ is a time evolution for a time $\delta t$, with a static unit charge inserted at site $n$.
The initial state $\Ket{h}$ is an eigenstate of the Hamiltonian with eigenvalue $E_h$, such that $\Bra{h} \tedag[ t_\z]$ = $\Bra{h} e^{i t_\z E_h}$ with $t_\z = \frac{\z}{2} \cdot \delta t$. Matrix elements in negative light-cone direction can be evaluated in an analogous way.
 
Using a second order Suzuki-Trotter decomposition~\cite{Trotter,OstmeyerTrotterization} of the time evolution operator, we find a time-dependent MPS (tMPS) approximation to the evolved state. In this work, the bond dimension of the MPS is fixed to be $D=80$ both for the initial state and after the truncation in each time-evolving block-decimation (TEBD)~\cite{TEBD} step. We follow the approach in~\cite{Schwinger_chiralCondensate} with a Trotter time step $\delta \tau = \frac{\delta t}{100}$ and truncate the electric fields to $\left|L_n\right| \le 10$.
Finally, \pdf can be computed by a discrete Fourier transform of the matrix elements.

\section{Numerical Results}
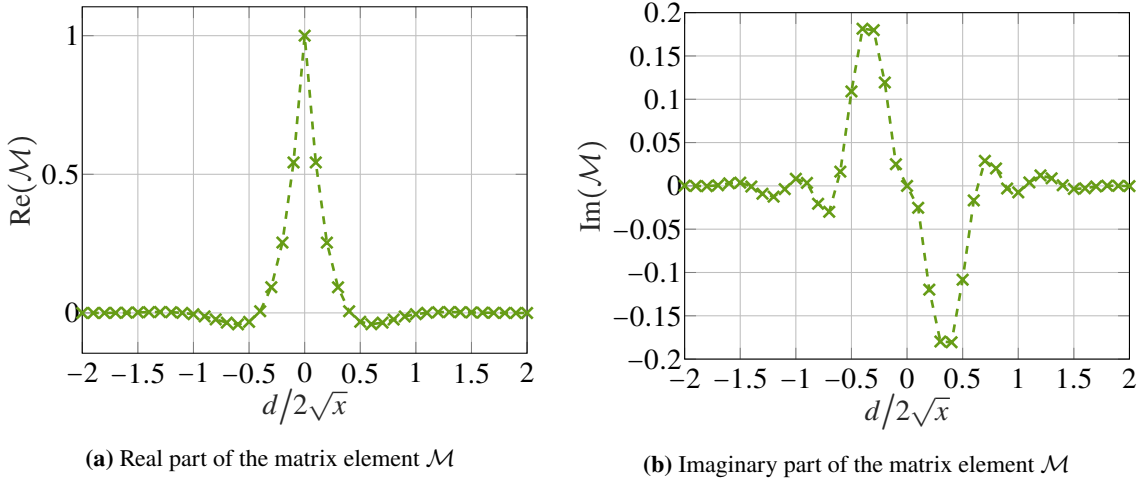
\begin{figure}[htb]
	\centering
	\def\figWidth{.77\columnwidth}
	\def\figHeight{.62\columnwidth}
	\begin{minipage}[t]{0.49\textwidth}
		\centering
%
%
\begin{tikzpicture}

\begin{axis}[%
width=\figWidth,
height=\figHeight,
at={(0\figWidth,0\figHeight)},
scale only axis,
xmin=-2,
xmax=2,
xlabel style={font=\color{white!15!black}},
xlabel={$d \big/ 2\sqrt{x}$},
ylabel style={font=\color{white!15!black}},
ylabel={$\Re(\cor)$},
axis background/.style={fill=white},
title style={font=\bfseries},
xmajorgrids,
ymajorgrids,
xtick={-2,-1.5,...,2},
ytick={-2,-1.8,...,2},
ylabel style={{rotate=-90}}, scaled ticks=false, tick label style={/pgf/number format/fixed},
xticklabel style={yshift=-.5ex},
yticklabel style={xshift=-.5ex}
]
\addplot [color=mycolor1, dashed, line width=1.0pt, mark size=3.0pt, mark=x, mark options={solid, mycolor1}]
  table[row sep=crcr]{%
-2.5	-1.36419331587525e-05\\
-2.4	-8.46138415495443e-06\\
-2.3	2.418541860943e-06\\
-2.2	3.40779693133069e-06\\
-2.1	3.7820028200009e-06\\
-2	1.90767201032802e-06\\
-1.9	-2.24908833953925e-06\\
-1.8	-1.70024334152563e-05\\
-1.7	-3.49079117263074e-05\\
-1.6	-3.14389080971051e-05\\
-1.5	1.6517181133051e-05\\
-1.4	6.79734123410782e-05\\
-1.3	4.94958016645851e-05\\
-1.2	-3.5676522320193e-05\\
-1.1	-5.56715347596803e-05\\
-1	0.00010554710005814\\
-0.9	0.00032419240547166\\
-0.8	0.000291294979703349\\
-0.7	-0.000110020670497212\\
-0.6	-0.000592223593478873\\
-0.5	-0.000795063186869401\\
-0.4	-0.000746426960976161\\
-0.3	-0.00073583953631013\\
-0.2	-0.000744308239459701\\
-0.1	-0.000329404076715889\\
0	1.00007103139128\\
0.1	-0.000264196539036893\\
0.2	-0.000899923487981902\\
0.3	-0.00111541294843565\\
0.4	-0.000909463680397679\\
0.5	-0.000456765894372067\\
0.6	-8.79874616230847e-05\\
0.7	2.44734898455917e-06\\
0.8	-6.5467398882752e-05\\
0.9	-5.319965940692e-05\\
1	8.84636394623438e-05\\
1.1	0.000201240560718138\\
1.2	0.000159285094041391\\
1.3	2.3293422473356e-05\\
1.4	-6.5624309888301e-05\\
1.5	-5.80361388864302e-05\\
1.6	-1.01316733672369e-05\\
1.7	1.34695153740436e-05\\
1.8	7.2345308009398e-06\\
1.9	-5.33669681697455e-06\\
2	-8.05683506900552e-06\\
2.1	-9.56563454376628e-06\\
2.2	-9.00558043166061e-06\\
2.3	-1.50726747289426e-06\\
2.4	1.0530005309695e-05\\
};

\end{axis}
\end{tikzpicture}%
		\subcaption{Real part of the matrix element \cor}
		\label{fig:CorrelatorExcit1_real}
	\end{minipage}
	\hfill
	\begin{minipage}[t]{0.49\textwidth}
		\centering
%
%
\begin{tikzpicture}

\begin{axis}[%
width=\figWidth,
height=\figHeight,
at={(0\figWidth,0\figHeight)},
scale only axis,
xmin=-2,
xmax=2,
xlabel style={font=\color{white!15!black}},
xlabel={$d \big/ 2\sqrt{x}$},
ymin=-0.08,
ymax=0.08,
ylabel style={font=\color{white!15!black}},
ylabel={$\Im(\cor)$},
axis background/.style={fill=white},
title style={font=\bfseries},
xmajorgrids,
ymajorgrids,
xtick={-2,-1.5,...,2},
ytick={-0.1,-0.08,...,0.1},
ylabel style={{rotate=-90}}, scaled ticks=false, tick label style={/pgf/number format/fixed},
xticklabel style={yshift=-.5ex},
yticklabel style={xshift=-.5ex}
]
\addplot [color=mycolor1, dashed, line width=1.0pt, mark size=3.0pt, mark=x, mark options={solid, mycolor1}]
  table[row sep=crcr]{%
-2.5	1.65605199804702e-05\\
-2.4	-3.5603619586322e-05\\
-2.3	-9.75894557658616e-05\\
-2.2	-4.0965279623184e-05\\
-2.1	0.000216142155733307\\
-2	0.000403801969122299\\
-1.9	-2.89795042551819e-05\\
-1.8	-0.00107477506284839\\
-1.7	-0.00130234201180252\\
-1.6	0.000891218011777823\\
-1.5	0.0039980854742385\\
-1.4	0.00283342674394436\\
-1.3	-0.00486433645745482\\
-1.2	-0.0109085096499339\\
-1.1	-0.0027671366815285\\
-1	0.0164827390088021\\
-0.9	0.0216895962162947\\
-0.8	-0.00494523915124984\\
-0.7	-0.0385133659465033\\
-0.6	-0.0299743695973506\\
-0.5	0.0251996187580832\\
-0.4	0.0620660047266804\\
-0.3	0.0234005636885907\\
-0.2	-0.0523670445301616\\
-0.1	-0.0679826574632349\\
0	-6.81455465727981e-16\\
0.1	0.067723364934783\\
0.2	0.0519579926104284\\
0.3	-0.0231721131894028\\
0.4	-0.0611427221258298\\
0.5	-0.0246641820991748\\
0.6	0.0293801088425749\\
0.7	0.0375020254327852\\
0.8	0.00471873175230213\\
0.9	-0.0210081796902878\\
1	-0.0158296317993629\\
1.1	0.00271393612407058\\
1.2	0.0104270739918848\\
1.3	0.00459037921663114\\
1.4	-0.00271181741579336\\
1.5	-0.00376689418673993\\
1.6	-0.000820920789991166\\
1.7	0.0012236023916955\\
1.8	0.000996128777018472\\
1.9	2.49907631217818e-05\\
2	-0.000372979285994842\\
2.1	-0.000196709970448677\\
2.2	3.69155705515422e-05\\
2.3	8.65978420661052e-05\\
2.4	4.26825642429005e-05\\
};

\end{axis}
\end{tikzpicture}%
		\subcaption{Imaginary part of the matrix element \cor}
		\label{fig:CorrelatorExcit1_imag}
	\end{minipage}
	\caption{Real and imaginary parts of the matrix element $\cor = \cor_{ee} - \cor_{eo} - \cor_{oe} + \cor_{oo}$ of the first excited state (vector boson) for different distances $d$ to the origin. Parameters: $\frac{m}{g} = 5.6419 \approx \frac{10}{\sqrt{\pi}}$, $x = 100$, $N=100$.  Dashed lines to guide the eye in all plots.} 
	\label{fig:CorrelatorExcit1}
\end{figure}

We calculate the matrix elements from the four contributions in \cref{eq:correlatorFull} for different distances to the origin. An example is given in \cref{fig:CorrelatorExcit1} for the first excited state (vector meson at rest) in the zero charge sector. The real part of the matrix element (\cref{fig:CorrelatorExcit1_real}) is found to be symmetric with respect to the axis $\z=0$, while the imaginary part (\cref{fig:CorrelatorExcit1_imag}) is antisymmetric. These symmetries make the PDF a real function as required.

\begin{figure}[htb]
	\centering
	\def\figWidth{0.91\textwidth}
	\def\figHeight{.62*0.5\textwidth}
%
%
\begin{tikzpicture}

\begin{axis}[%
width=\figWidth,
height=\figHeight,
at={(0\figWidth,0\figHeight)},
scale only axis,
xmin=-3.14159265358979,
xmax=3.14160265358979,
xtick={-6.283185307179586, -5.497787143782138, -4.712388980384690, -3.926990816987241, -3.141592653589793, -2.356194490192345, -1.570796326794897, -0.785398163397448, 0, 0.785398163397448, 1.570796326794897, 2.356194490192345, 3.141592653589793, 3.926990816987241, 4.712388980384690, 5.497787143782138, 6.283185307179586},
xticklabels={{$\text{-2}\pi$},{$\text{-7}\pi\text{/4}$},{$\text{-3}\pi\text{/2}$},{$\text{-5}\pi\text{/4}$},{$\text{-}\pi$},{$\text{-3}\pi\text{/4}$},{$\text{-}\pi\text{/2}$},{$\text{-}\pi\text{/4}$},{0},{$\text{-}\pi\text{/4}$},{$\pi\text{/2}$},{$\text{-3}\pi\text{/4}$},{$\pi$},{$\text{-5}\pi\text{/4}$},{$\text{3}\pi\text{/2}$},{$\text{-7}\pi\text{/4}$},{$\text{2}\pi$}},
xlabel style={font=\color{white!15!black}},
xlabel={$k$},
ymin=0.3,
ymax=1.7,
ylabel style={font=\color{white!15!black}},
ylabel={$\sum_\z \cor(\z) e^{-i k \z}$},
axis background/.style={fill=white},
title style={font=\bfseries},
xmajorgrids,
ymajorgrids,
legend style={at={(0.02,0.94)}, anchor=north west, legend cell align=left, align=left, draw=white!15!black},
ylabel style={{rotate=-90}}, scaled ticks=false, tick label style={/pgf/number format/fixed},
xticklabel style={yshift=-.5ex},
yticklabel style={xshift=-.5ex}
]
\addplot [color=black, dashed, line width=1.0pt, mark size=3.0pt, mark=o, mark options={solid, black}]
  table[row sep=crcr]{%
-3.14159265358979	0.999888585244044\\
-3.0159289474462	1.00007765990475\\
-2.89026524130261	1.00024973013717\\
-2.76460153515902	1.00045961339794\\
-2.63893782901543	1.00064873256817\\
-2.51327412287183	1.00088817809767\\
-2.38761041672824	1.0011072967789\\
-2.26194671058465	1.0013892117117\\
-2.13628300444106	1.00165815604818\\
-2.01061929829747	1.0020035582076\\
-1.88495559215388	1.00235400697524\\
-1.75929188601028	1.00279977864721\\
-1.63362817986669	1.00328561335487\\
-1.5079644737231	1.00389981476552\\
-1.38230076757951	1.00461810459745\\
-1.25663706143592	1.00552839566817\\
-1.13097335529233	1.0066502132097\\
-1.00530964914873	1.0080510590592\\
-0.879645943005142	1.00969311107776\\
-0.75398223686155	1.01117174188661\\
-0.628318530717959	1.01105735570003\\
-0.502654824574367	1.00690534814496\\
-0.376991118430775	0.999910871433594\\
-0.251327412287183	0.994889813545449\\
-0.125663706143592	0.993036552788584\\
0	0.992842062242677\\
0.125663706143592	0.993284200578247\\
0.251327412287183	0.993851759808438\\
0.376991118430775	0.9944460821078\\
0.502654824574367	0.994957083396388\\
0.628318530717959	0.995442684247055\\
0.75398223686155	0.995839167412539\\
0.879645943005142	0.996224073095778\\
1.00530964914873	0.996534288342028\\
1.13097335529233	0.996847634433446\\
1.25663706143592	0.997099358036327\\
1.38230076757951	0.997363055165932\\
1.5079644737231	0.997576042213409\\
1.63362817986669	0.997804835808485\\
1.75929188601028	0.997992642856224\\
1.88495559215388	0.998196613119736\\
2.01061929829747	0.99836894887934\\
2.13628300444106	0.998555118999388\\
2.26194671058465	0.998719511112092\\
2.38761041672824	0.998893169028803\\
2.51327412287183	0.999055702266402\\
2.63893782901543	0.999221497275052\\
2.76460153515902	0.999387472826892\\
2.89026524130261	0.999549877000898\\
3.0159289474462	0.999724616796051\\
};
\addlegendentry{Ground state}

\addplot [color=mycolor1, dashed, line width=1.0pt, mark size=3.0pt, mark=x, mark options={solid, mycolor1}]
  table[row sep=crcr]{%
-3.14159265358979	0.999979439089245\\
-3.0159289474462	1.00013717605695\\
-2.89026524130261	1.00034609347038\\
-2.76460153515902	1.00052288680048\\
-2.63893782901543	1.00075915437183\\
-2.51327412287183	1.00095376381637\\
-2.38761041672824	1.0012307383852\\
-2.26194671058465	1.00145910575027\\
-2.13628300444106	1.00180652983259\\
-2.01061929829747	1.00204899271871\\
-1.88495559215388	1.00230197157251\\
-1.75929188601028	1.00113167186752\\
-1.63362817986669	0.991046735553821\\
-1.5079644737231	0.937005745519554\\
-1.38230076757951	0.769070930792351\\
-1.25663706143592	0.514237953995369\\
-1.13097335529233	0.429460821748649\\
-1.00530964914873	0.632315012355314\\
-0.879645943005142	0.87167474896224\\
-0.75398223686155	0.980542245012619\\
-0.628318530717959	1.00662016746111\\
-0.502654824574367	1.00666794290499\\
-0.376991118430775	1.00026747098999\\
-0.251327412287183	0.995284568060636\\
-0.125663706143592	0.993480197799945\\
0	0.993171515215549\\
0.125663706143592	0.993612081850402\\
0.251327412287183	0.99411840227543\\
0.376991118430775	0.994765817243988\\
0.502654824574367	0.995793078824766\\
0.628318530717959	1.00041211489494\\
0.75398223686155	1.02673665145165\\
0.879645943005142	1.13412934614957\\
1.00530964914873	1.37169993360153\\
1.13097335529233	1.57287890685223\\
1.25663706143592	1.48728619156647\\
1.38230076757951	1.23273427311389\\
1.5079644737231	1.0648289578279\\
1.63362817986669	1.01048489267887\\
1.75929188601028	0.999915509134247\\
1.88495559215388	0.998515721681689\\
2.01061929829747	0.998471096123573\\
2.13628300444106	0.998631564392105\\
2.26194671058465	0.998778909596921\\
2.38761041672824	0.99896727234814\\
2.51327412287183	0.999110090384194\\
2.63893782901543	0.999302420539531\\
2.76460153515902	0.999439321803253\\
2.89026524130261	0.999637421335823\\
3.0159289474462	0.999778013788821\\
};
\addlegendentry{First excited state}

\end{axis}
\end{tikzpicture}%
	\caption{Fourier transform of the matrix elements \cor for the ground state (vacuum) and the first excited state (vector boson). Parameters as in \cref{fig:CorrelatorExcit1}.}
	\label{fig:FourierTransoformVacExcit}
\end{figure}
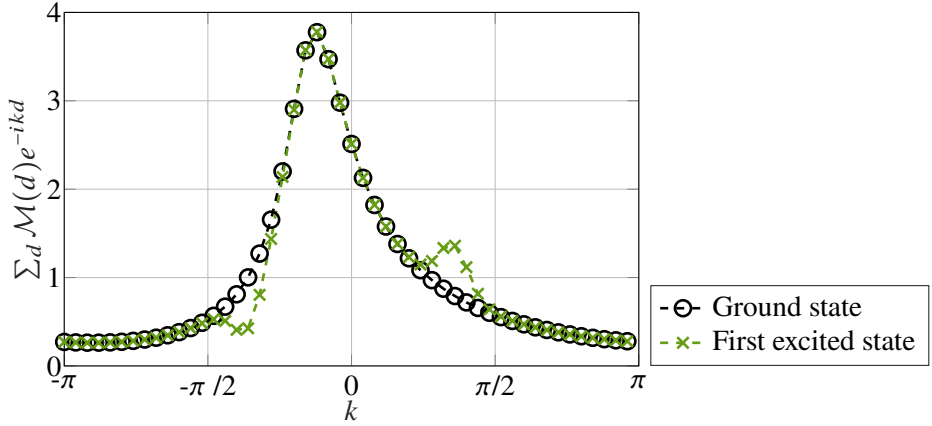

The PDF is related to the Fourier transform of the matrix elements, which is shown in \cref{fig:FourierTransoformVacExcit} for the ground state (vacuum) and the first excited state (vector meson). The vacuum matrix element is a background that needs to be subtracted to obtain physical PDFs~\cite{collins_2011}. We find that the excited state coincides with the vacuum contribution for most of Fourier modes $k$, but a region with positive $k$ shows some excess, while a region for negative $k$ shows a lower value than the vacuum.

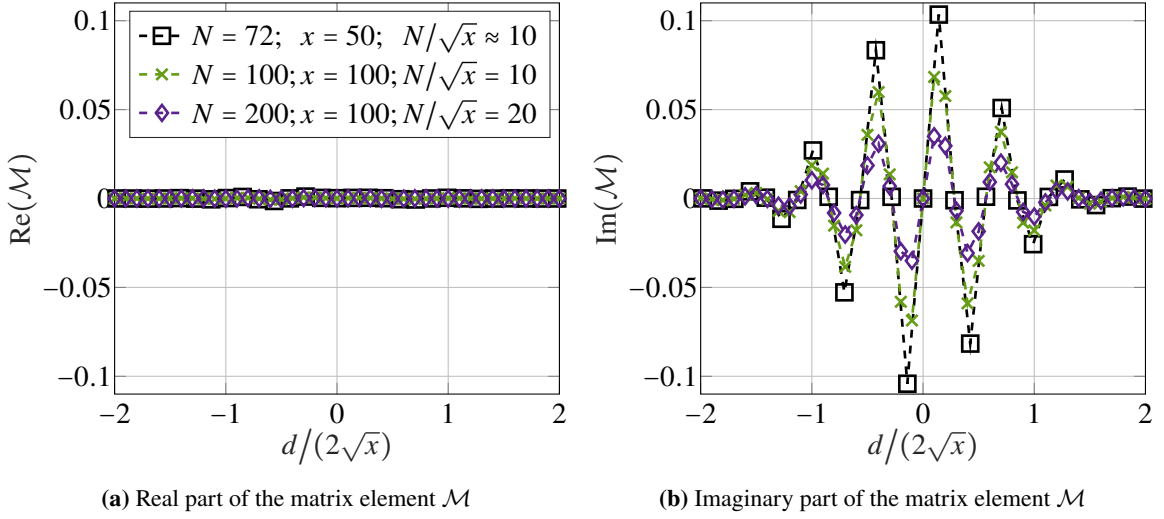
\begin{figure}[htb]
	\centering
	\def\figWidth{.79\columnwidth}
	\def\figHeight{.7\columnwidth}
	\begin{minipage}[t]{0.49\textwidth}
		\centering
%
%
\begin{tikzpicture}

\begin{axis}[%
width=\figWidth,
height=\figHeight,
at={(0\figWidth,0\figHeight)},
scale only axis,
xmin=-2,
xmax=2,
xlabel style={font=\color{white!15!black}},
xlabel={$d \big/ (2\sqrt{x})$},
ymin=-8,
ymax=8,
ylabel style={font=\color{white!15!black}},
ylabel={$\Re(\cor) \cdot N$},
axis background/.style={fill=white},
xmajorgrids,
ymajorgrids,
legend style={legend cell align=left, align=left, draw=white!15!black},
ylabel style={{rotate=-90}}, scaled ticks=false, tick label style={/pgf/number format/fixed},
xticklabel style={yshift=-.5ex},
yticklabel style={xshift=-.5ex}
]
\addplot [color=black, dashed, line width=1.0pt, mark size=3.0pt, mark=square, mark options={solid, black}]
  table[row sep=crcr]{%
-2.54558441227157	-0.000647763276579563\\
-2.40416305603426	-0.000601961188247774\\
-2.26274169979695	-0.000213401746687541\\
-2.12132034355964	0.000439901003402492\\
-1.97989898732233	0.00125139983797117\\
-1.83847763108502	-0.000798817804447348\\
-1.69705627484771	-0.00446271544903352\\
-1.5556349186104	0.0021338999375531\\
-1.41421356237309	0.0119125812851118\\
-1.27279220613579	-0.00575832056599643\\
-1.13137084989848	-0.0253635780583294\\
-0.989949493661166	0.0122782640660674\\
-0.848528137423857	0.0416939421972767\\
-0.707106781186547	-0.0196835647879247\\
-0.565685424949238	-0.0508478240457038\\
-0.424264068711929	0.0192021719140802\\
-0.282842712474619	0.0401545251073537\\
-0.14142135623731	0.000725782217953608\\
0	0.0134901350578644\\
0.14142135623731	0.001464904717972\\
0.282842712474619	-0.00915697634767737\\
0.424264068711929	0.013065007880166\\
0.565685424949238	0.0195449959556597\\
0.707106781186547	-0.00904027932448342\\
0.848528137423857	-0.0179784645520826\\
0.989949493661166	0.00315146082679986\\
1.13137084989848	0.0116956012675501\\
1.27279220613579	-0.000516843877392023\\
1.41421356237309	-0.00618018245791197\\
1.5556349186104	-0.00032172167580298\\
1.69705627484771	0.00254532976729459\\
1.83847763108502	0.000564427490212974\\
1.97989898732233	-0.000638466519721826\\
2.12132034355964	-0.000819855638139828\\
2.26274169979695	-0.000493024518981966\\
2.40416305603426	0.00121500813253197\\
};
\addlegendentry{$N=72; x=50; N/\sqrt{x} \approx 10$}

\addplot [color=mycolor1, dashed, line width=1.0pt, mark size=3.0pt, mark=x, mark options={solid, mycolor1}]
  table[row sep=crcr]{%
-2.5	-0.00141551021844336\\
-2.4	-0.000972984766781817\\
-2.3	8.13480614257705e-05\\
-2.2	0.00018309002450663\\
-2.1	0.00030631085235157\\
-2	0.000318381980613351\\
-1.9	0.000226919764915044\\
-1.8	-0.000833501164387671\\
-1.7	-0.00222739133333305\\
-1.6	-0.00170261212766509\\
-1.5	0.0027635110098439\\
-1.4	0.00675965507698681\\
-1.3	0.0027184212170841\\
-1.2	-0.00898563313675425\\
-1.1	-0.0146185262840406\\
-1	-0.00133701573657147\\
-0.9	0.0204724902000401\\
-0.8	0.0224319991680609\\
-0.7	-0.0047597686571661\\
-0.6	-0.0312767233925373\\
-0.5	-0.022964703792457\\
-0.4	0.0108329003833119\\
-0.3	0.0288034953974061\\
-0.2	0.0154742210825773\\
-0.1	-0.00131621516923985\\
0	0.00710313912837535\\
0.1	0.00520453859712688\\
0.2	-8.73037702699975e-05\\
0.3	-0.0091538458155056\\
0.4	-0.00547077155689185\\
0.5	0.0108650254527789\\
0.6	0.0191468896705296\\
0.7	0.00648703367329495\\
0.8	-0.0132442368744653\\
0.9	-0.0172667186703582\\
1	-0.00304536451578812\\
1.1	0.0110726605131706\\
1.2	0.0105106168207705\\
1.3	9.81696407696409e-05\\
1.4	-0.00660025421160888\\
1.5	-0.00469182991851791\\
1.6	0.0004284205340838\\
1.7	0.00261103380559474\\
1.8	0.00159051766424727\\
1.9	-8.31967764055291e-05\\
2	-0.000679260055807222\\
2.1	-0.00102691116968081\\
2.2	-0.00106729501607247\\
2.3	-0.000256383997092906\\
2.4	0.00153846569512589\\
};
\addlegendentry{$N=100; x=100; N/\sqrt{x}=10$}

\addplot [color=mycolor2, dashed, line width=1.0pt, mark size=3.0pt, mark=diamond, mark options={solid, mycolor2}]
  table[row sep=crcr]{%
-5	1.0603982490095e-06\\
-4.9	6.33847975321596e-05\\
-4.8	-0.000278720901779781\\
-4.7	0.000238293332346474\\
-4.6	-0.000262884087096187\\
-4.5	-7.64467525080615e-05\\
-4.4	-0.000376335851128374\\
-4.3	0.000552256423697796\\
-4.2	0.000183360898806216\\
-4.1	3.0335499754491e-05\\
-4	0.000376063020068584\\
-3.9	-0.000338834771515555\\
-3.8	-0.000370873642497319\\
-3.7	0.000104289935437197\\
-3.6	-1.35435274555215e-05\\
-3.5	0.000159888480595646\\
-3.4	0.00014006898114092\\
-3.3	-7.70028881476181e-05\\
-3.2	-1.38773872780011e-06\\
-3.1	8.49472211614884e-05\\
-3	5.55428539293636e-05\\
-2.9	0.000248379182963452\\
-2.8	-0.000140509390673692\\
-2.7	-0.000129025378076022\\
-2.6	-0.000169606750493198\\
-2.5	-0.000401318972011783\\
-2.4	0.000307244049004226\\
-2.3	0.000451093239499464\\
-2.2	-0.000119774833459896\\
-2.1	-0.00114354682451428\\
-2	-0.00208188086324602\\
-1.9	-0.000175167665925641\\
-1.8	0.00322295849892172\\
-1.7	0.00346844065281158\\
-1.6	-0.00328606025498261\\
-1.5	-0.00968377124040343\\
-1.4	-0.00389576410039719\\
-1.3	0.0118380399704743\\
-1.2	0.0172389523965325\\
-1.1	-0.00181922604983878\\
-1	-0.0257818226536078\\
-0.9	-0.0191944931632889\\
-0.8	0.0165543512686082\\
-0.7	0.0352217931953418\\
-0.6	0.00764983060964353\\
-0.5	-0.0304237677029353\\
-0.4	-0.0264618542411041\\
-0.3	0.010941222583555\\
-0.2	0.0271196504260764\\
-0.1	0.00794439675210157\\
0	0.00191508104344962\\
0.1	0.00965868561624461\\
0.2	0.0229929873968189\\
0.3	0.000905850647691331\\
0.4	-0.0307243481073201\\
0.5	-0.0213599138539955\\
0.6	0.0210921179774992\\
0.7	0.0381256816956235\\
0.8	0.00686762721879369\\
0.9	-0.0294220189256341\\
1	-0.026060975056488\\
1.1	0.00534862922379706\\
1.2	0.022529264238659\\
1.3	0.0108070658027555\\
1.4	-0.00786465665429886\\
1.5	-0.0116549536728441\\
1.6	-0.0025290637490004\\
1.7	0.00467035684018909\\
1.8	0.0038293249278256\\
1.9	-0.00037545758234224\\
2	-0.00203476704563022\\
2.1	-0.0011699409044637\\
2.2	0.000294464259748922\\
2.3	0.000711172172911922\\
2.4	0.000124499209644588\\
2.5	-7.71427410030103e-05\\
2.6	-0.000333473741142771\\
2.7	-0.000484081455529708\\
2.8	-0.000164769934779532\\
2.9	8.25029441063733e-05\\
3	-0.000152686442517363\\
3.1	-0.000158033141468039\\
3.2	7.3093647553188e-05\\
3.3	0.000122104144363689\\
3.4	0.000273625234019138\\
3.5	0.000384071539131897\\
3.6	-3.28582446322454e-05\\
3.7	-0.000327215663060299\\
3.8	-0.000198185039308939\\
3.9	0.000195556150497421\\
4	6.19716862328263e-05\\
4.1	-0.000356436492740696\\
4.2	0.000168486107968326\\
4.3	8.24600637654256e-05\\
4.4	-0.00020172641861722\\
4.5	0.000422974089215312\\
4.6	-0.000779370360420611\\
4.7	-2.96863755249458e-05\\
4.8	0.000710196801023836\\
4.9	9.9799200679727e-06\\
};
\addlegendentry{$N=200; x=100; N/\sqrt{x}=20$}

\end{axis}
\end{tikzpicture}%
		\subcaption{Real part of the matrix element \cor}
		\label{fig:CorrelatorSubtracted_real}
	\end{minipage}
	\hfill
	\begin{minipage}[t]{0.49\textwidth}
		\centering
%
%
\begin{tikzpicture}

\begin{axis}[%
width=\figWidth,
height=\figHeight,
at={(0\figWidth,0\figHeight)},
scale only axis,
xmin=-2,
xmax=2,
xlabel style={font=\color{white!15!black}},
xlabel={$d \big/ (2\sqrt{x})$},
ymin=-8,
ymax=8,
ylabel style={font=\color{white!15!black}},
ylabel={$\Im(\cor) \cdot N$},
axis background/.style={fill=white},
xmajorgrids,
ymajorgrids,
ylabel style={{rotate=-90}}, scaled ticks=false, tick label style={/pgf/number format/fixed},
xticklabel style={yshift=-.5ex},
yticklabel style={xshift=-.5ex}
]
\addplot [color=black, dashed, line width=1.0pt, mark size=3.0pt, mark=square, mark options={solid, black}]
  table[row sep=crcr]{%
-2.54558441227157	0.00232349876776941\\
-2.40416305603426	-0.00345354110154798\\
-2.26274169979695	-0.0101976546600339\\
-2.12132034355964	0.0160797186650944\\
-1.97989898732233	0.0401164936734924\\
-1.83847763108502	-0.0682508951108527\\
-1.69705627484771	-0.12994878330765\\
-1.5556349186104	0.246592079163153\\
-1.41421356237309	0.341851675879871\\
-1.27279220613579	-0.741860880089953\\
-1.13137084989848	-0.716902551227659\\
-0.989949493661166	1.83400086829115\\
-0.848528137423857	1.1606823601539\\
-0.707106781186547	-3.68041548219369\\
-0.565685424949238	-1.35963321982575\\
-0.424264068711929	5.91530007528865\\
-0.282842712474619	0.961744664135685\\
-0.14142135623731	-7.50057332600353\\
0	-1.13619982776725e-13\\
0.14142135623731	7.44724807321515\\
0.282842712474619	-0.951414654359895\\
0.424264068711929	-5.78733699474754\\
0.565685424949238	1.3313972483098\\
0.707106781186547	3.54623799973663\\
0.848528137423857	-1.12332787850394\\
0.989949493661166	-1.73891142086817\\
1.13137084989848	0.684078803205602\\
1.27279220613579	0.691018272134207\\
1.41421356237309	-0.320845008345253\\
1.5556349186104	-0.225229221993908\\
1.69705627484771	0.119601667378312\\
1.83847763108502	0.0611029894386152\\
1.97989898732233	-0.0360534748601554\\
2.12132034355964	-0.0142243976618189\\
2.26274169979695	0.00898768528258019\\
2.40416305603426	0.00340268506527603\\
};

\addplot [color=mycolor1, dashed, line width=1.0pt, mark size=3.0pt, mark=x, mark options={solid, mycolor1}]
  table[row sep=crcr]{%
-2.5	0.00159851924837943\\
-2.4	-0.00358030790988884\\
-2.3	-0.00968975355920187\\
-2.2	-0.00390952129160433\\
-2.1	0.0219438498094239\\
-2	0.0408347656203673\\
-1.9	-0.00241731411752107\\
-1.8	-0.107174392111173\\
-1.7	-0.13041846176761\\
-1.6	0.0880820593422123\\
-1.5	0.397603356909738\\
-1.4	0.279913769386882\\
-1.3	-0.49063515924158\\
-1.2	-1.09460882336512\\
-1.1	-0.277907570990493\\
-1	1.65252271830007\\
-0.9	2.18167296249031\\
-0.8	-0.471405203896476\\
-0.7	-3.81877354927859\\
-0.6	-2.96141935181246\\
-0.5	2.546670899693\\
-0.4	6.20424054188569\\
-0.3	2.28467017140179\\
-0.2	-5.36589456109471\\
-0.1	-7.00646530672796\\
0	-2.65623688225602e-14\\
0.1	6.98053605388359\\
0.2	5.32498936912277\\
0.3	-2.2618251214808\\
0.4	-6.11191228179803\\
0.5	-2.4931272337924\\
0.6	2.9019932763187\\
0.7	3.71763949751341\\
0.8	0.448754463946059\\
0.9	-2.11353130400834\\
1	-1.58721198575306\\
1.1	0.272587511236122\\
1.2	1.04646527548781\\
1.3	0.463239528627872\\
1.4	-0.267752853625209\\
1.5	-0.374484469643035\\
1.6	-0.0810526592685681\\
1.7	0.122545169049254\\
1.8	0.0993106599884909\\
1.9	0.00201914245895223\\
2	-0.037753432459906\\
2.1	-0.0200038602628575\\
2.2	0.00350898188992918\\
2.3	0.00869864634037236\\
2.4	0.00348253528608653\\
};

\addplot [color=mycolor2, dashed, line width=1.0pt, mark size=3.0pt, mark=diamond, mark options={solid, mycolor2}]
  table[row sep=crcr]{%
-5	0.00010593815057825\\
-4.9	-0.00017956040292716\\
-4.8	-0.000496597537189103\\
-4.7	-9.42514328962806e-05\\
-4.6	0.000470208999392469\\
-4.5	0.000271455302681754\\
-4.4	-0.000281811513193069\\
-4.3	-8.1574301546814e-05\\
-4.2	0.000175035202456838\\
-4.1	-2.16894906969626e-06\\
-4	0.000291999596514442\\
-3.9	-9.36685224571123e-05\\
-3.8	-0.00054073747572854\\
-3.7	-0.000152439410891828\\
-3.6	0.00049158405440607\\
-3.5	0.000152179307276066\\
-3.4	6.32734161121219e-05\\
-3.3	-6.98247729152222e-05\\
-3.2	-6.85603648482305e-05\\
-3.1	0.000286604124952279\\
-3	4.0952072380142e-05\\
-2.9	-0.000803263456505121\\
-2.8	-0.00129681158383738\\
-2.7	0.000387190911718462\\
-2.6	0.00319275024884893\\
-2.5	0.00374581511705487\\
-2.4	-0.00476666688469038\\
-2.3	-0.0169222572964324\\
-2.2	-0.00928511547076355\\
-2.1	0.0309990496044911\\
-2	0.0628293492186606\\
-1.9	0.0022090804779829\\
-1.8	-0.14363620011828\\
-1.7	-0.176643437513892\\
-1.6	0.105013181700809\\
-1.5	0.490940881241583\\
-1.4	0.342635708326925\\
-1.3	-0.571200382313737\\
-1.2	-1.25573770364099\\
-1.1	-0.31148071061128\\
-1	1.82541874768475\\
-0.9	2.36202652853287\\
-0.8	-0.519230310088271\\
-0.7	-4.04032417181655\\
-0.6	-3.07852852036701\\
-0.5	2.66300245345689\\
-0.4	6.38830210825548\\
-0.3	2.32093676098332\\
-0.2	-5.49423166668791\\
-0.1	-7.14659325583208\\
0	-4.30813513619118e-13\\
0.1	7.13958988411782\\
0.2	5.48315603592565\\
0.3	-2.31477074190084\\
0.4	-6.36316729197248\\
0.5	-2.64829699010912\\
0.6	3.06226695994344\\
0.7	4.01229260199649\\
0.8	0.512775355981506\\
0.9	-2.34295515881971\\
1	-1.80682841793288\\
1.1	0.310138721587911\\
1.2	1.24175121827551\\
1.3	0.562988439118851\\
1.4	-0.339374059079218\\
1.5	-0.483997954178062\\
1.6	-0.102513585024261\\
1.7	0.174464560660641\\
1.8	0.141144792849353\\
1.9	-0.0024382148280735\\
2	-0.0612962448814584\\
2.1	-0.0301456054635801\\
2.2	0.00933104447693319\\
2.3	0.016114436415709\\
2.4	0.00461948147304556\\
2.5	-0.00358249059734011\\
2.6	-0.00339176731896711\\
2.7	-0.000159404894661839\\
2.8	0.00107026951473704\\
2.9	0.000530337402505208\\
3	0.000101686598415537\\
3.1	-1.42675691881595e-05\\
3.2	0.000114658498270442\\
3.3	0.000130562992020143\\
3.4	0.000184359572389091\\
3.5	-2.69101415535698e-05\\
3.6	-0.000242023841112678\\
3.7	5.54673331757987e-06\\
3.8	-9.07249631668224e-05\\
3.9	-7.71400337787182e-05\\
4	0.000715687771760531\\
4.1	-0.000528868006235576\\
4.2	0.000428758166505527\\
4.3	-0.00040967393219098\\
4.4	-5.55192567558276e-05\\
4.5	0.000130369810739723\\
4.6	0.000329568221098659\\
4.7	0.000210746575280535\\
4.8	-0.000198855094217916\\
4.9	0.000891471184072985\\
};

\end{axis}
\end{tikzpicture}%
		\subcaption{Imaginary part of the matrix element \cor}
		\label{fig:CorrelatorSubtracted_imag}
	\end{minipage}
	\caption{Real and imaginary parts of the matrix element \cor. For each data point, the difference between the matrix elements of the first excited state (vector boson) and the ground state (vacuum) is taken. Parameters: $\frac{m}{g} = 5.6419 \approx \frac{10}{\sqrt{\pi}}$; different symbols for different lattice spacings and physical volumes.}
	\label{fig:CorrelatorSubtracted}
\end{figure}

In \cref{fig:CorrelatorSubtracted}, we plot the matrix element of the first excited state with the vacuum contribution subtracted for each point. Different lattice spacings and physical volumes are considered. The real part in \cref{fig:CorrelatorSubtracted_real} vanishes and the imaginary part follows a damped oscillation. The periodicity and form of the matrix element do not change significantly once the lattice spacing is chosen small enough and the physical volume is large enough. This corresponds to sufficiently large $x$ and $\frac{L}{x}$. Only the amplitude of the lattice matrix elements scales with $N$, and the resolution varies with $x$.

\begin{figure}[H]
	\centering
	\def\figWidth{.8\columnwidth}
	\def\figHeight{.69\columnwidth}
	\begin{minipage}[t]{0.49\textwidth}
		\centering
%
%
\begin{tikzpicture}

\begin{axis}[%
width=\figWidth,
height=\figHeight,
at={(0\figWidth,0\figHeight)},
scale only axis,
xmin=-1,
xmax=1,
xlabel style={font=\color{white!15!black}},
xlabel={\bjorx},
ymin=-6,
ymax=6,
ylabel style={font=\color{white!15!black}},
ylabel={\pdf},
axis background/.style={fill=white},
xmajorgrids,
ymajorgrids,
legend style={at={(0.5,-0.16)}, anchor=north, legend cell align=left, align=left, draw=white!15!black},
ylabel style={{rotate=-90}}, scaled ticks=false, tick label style={/pgf/number format/fixed},
xticklabel style={yshift=-.5ex},
yticklabel style={xshift=-.5ex}
]
\addplot [color=black, dashed, line width=1.0pt, mark size=3.0pt, mark=square, mark options={solid, black}]
  table[row sep=crcr]{%
-0.958240858194233	0.00125316225259551\\
-0.90500525496122	0.00175723225525804\\
-0.851769651728208	0.000999136813216543\\
-0.798534048495194	-0.00145929207744076\\
-0.745298445262181	-0.024914065550846\\
-0.692062842029169	-0.168894016210805\\
-0.638827238796155	-0.824563293745019\\
-0.585591635563143	-2.58776917416379\\
-0.53235603233013	-4.83993341223079\\
-0.479120429097117	-5.16777092876046\\
-0.425884825864104	-3.11647326949234\\
-0.372649222631091	-1.08960291195442\\
-0.319413619398078	-0.236856285322218\\
-0.266178016165065	-0.0334957520542558\\
-0.212942412932052	0.000931038251701831\\
-0.159706809699039	0.00618830628476112\\
-0.106471206466026	0.00671511254330875\\
-0.053235603233013	0.00644666932828833\\
0	0.00560996440246769\\
0.053235603233013	0.00531976618218447\\
0.106471206466026	0.00475103838250226\\
0.159706809699039	0.00535910801484166\\
0.212942412932052	0.009530386432092\\
0.266178016165065	0.0419680270599051\\
0.319413619398078	0.239136317367048\\
0.372649222631091	1.07677014245245\\
0.425884825864104	3.0873766866751\\
0.479120429097117	5.14311244226516\\
0.53235603233013	4.84220540285614\\
0.585591635563143	2.60861451794442\\
0.638827238796155	0.842532679439772\\
0.692062842029169	0.179507990629155\\
0.745298445262181	0.0300872206346103\\
0.798534048495194	0.00557343654808411\\
0.851769651728208	0.00176674556597194\\
0.90500525496122	0.00157096610663058\\
};
\addlegendentry{$N=72; x=50; N/\sqrt{x} \approx 10$}

\addplot [color=mycolor1, dashed, line width=1.0pt, mark size=3.0pt, mark=x, mark options={solid, mycolor1}]
  table[row sep=crcr]{%
-1.35466027924186	0.000838345290258488\\
-1.30047386807218	0.00054917968285069\\
-1.24628745690251	0.000889183571367978\\
-1.19210104573284	0.000583849356016363\\
-1.13791463456316	0.00101890678197238\\
-1.08372822339349	0.000605186035397446\\
-1.02954181222381	0.00113904578321627\\
-0.975355401054139	0.000644940650800777\\
-0.921168989884464	0.00136910510592388\\
-0.86698257871479	0.000419242667425688\\
-0.812796167545115	-0.000480151772440666\\
-0.758609756375441	-0.015392298029013\\
-0.704423345205767	-0.112933090943429\\
-0.650236934036092	-0.617258716733466\\
-0.596050522866418	-2.17348933727618\\
-0.541864111696744	-4.53333622828815\\
-0.487677700527069	-5.32596068831443\\
-0.433491289357395	-3.46706894397698\\
-0.379304878187721	-1.27355142309889\\
-0.325118467018046	-0.282630794444762\\
-0.270932055848372	-0.0409437361059881\\
-0.216745644678697	-0.00219063446759925\\
-0.162559233509023	0.00329048878393862\\
-0.108372822339349	0.00364256006871212\\
-0.0541864111696744	0.00409369251242122\\
0	0.00303999624407561\\
0.0541864111696744	0.0030254935239069\\
0.108372822339349	0.0024604182232816\\
0.162559233509023	0.00295032582235042\\
0.216745644678697	0.00771406899194706\\
0.270932055848372	0.0458549527511707\\
0.325118467018046	0.285103620743221\\
0.379304878187721	1.27250790444457\\
0.433491289357395	3.46180561843028\\
0.487677700527069	5.31527425404735\\
0.541864111696744	4.52315278820756\\
0.596050522866418	2.1718657211947\\
0.650236934036092	0.620569937764478\\
0.704423345205767	0.11700402920832\\
0.758609756375441	0.0177430672793768\\
0.812796167545115	0.00294454416766559\\
0.86698257871479	0.000942554065015977\\
0.921168989884464	0.000705392653500077\\
0.975355401054139	0.00054809391825387\\
1.02954181222381	0.000683781392219238\\
1.08372822339349	0.000501861228832388\\
1.13791463456316	0.000746711792978284\\
1.19210104573284	0.000478431540695429\\
1.24628745690251	0.000807807096234054\\
1.30047386807218	0.000492715716145149\\
};
\addlegendentry{$N=100; x=100; N/\sqrt{x}=10$}

\addplot [color=mycolor2, dashed, line width=1.0pt, mark size=3.0pt, mark=diamond, mark options={solid, mycolor2}]
  table[row sep=crcr]{%
-1.35501084907905	0.000504382826829447\\
-1.32791063209747	0.000949743032941561\\
-1.30081041511589	0.000492355437304233\\
-1.27371019813431	0.000946029596543135\\
-1.24660998115273	0.000592748751396311\\
-1.21950976417115	0.000851805926346864\\
-1.19240954718957	0.000728738143349855\\
-1.16530933020799	0.000617968856766077\\
-1.13820911322641	0.00128924352251468\\
-1.11110889624482	0.000456174505728412\\
-1.08400867926324	0.00119993580077589\\
-1.05690846228166	0.000449387216760851\\
-1.02980824530008	0.00117882647117408\\
-1.0027080283185	0.000882968190780319\\
-0.975607811336919	0.000936239084732299\\
-0.948507594355338	0.00111479406091308\\
-0.921407377373757	0.00106602906061547\\
-0.894307160392176	0.0009197069274708\\
-0.867206943410595	0.000841375673245757\\
-0.840106726429014	0.000730701522184843\\
-0.813006509447433	-0.000463865998529926\\
-0.785906292465852	-0.00343369348327526\\
-0.758806075484271	-0.0116038817496895\\
-0.73170585850269	-0.0348131552029685\\
-0.704605641521108	-0.0951012988558828\\
-0.677505424539527	-0.241739562667532\\
-0.650405207557946	-0.562030631978465\\
-0.623304990576365	-1.17174265945886\\
-0.596204773594784	-2.15530740197902\\
-0.569104556613203	-3.44726635690626\\
-0.542004339631622	-4.74861359534167\\
-0.514904122650041	-5.59728008967756\\
-0.48780390566846	-5.62421157624972\\
-0.460703688686879	-4.81756023727564\\
-0.433603471705298	-3.5280561640236\\
-0.406503254723716	-2.22229825749923\\
-0.379403037742135	-1.21545317805714\\
-0.352302820760554	-0.585524674789333\\
-0.325202603778973	-0.251424398292007\\
-0.298102386797392	-0.0988202191688446\\
-0.271002169815811	-0.0344037881798868\\
-0.24390195283423	-0.00965808216930832\\
-0.216801735852649	-0.000646399390827103\\
-0.189701518871068	0.00242728933192848\\
-0.162601301889487	0.00364264216021489\\
-0.135501084907905	0.00353467545926265\\
-0.108400867926324	0.0039612385290523\\
-0.0813006509447433	0.0045080799594552\\
-0.0542004339631622	0.0043490455644903\\
-0.0271002169815811	0.00341255670216319\\
0	0.0030867182209293\\
0.0271002169815811	0.00318555191696442\\
0.0542004339631622	0.00305352347352283\\
0.0813006509447433	0.00278548880744603\\
0.108400867926324	0.00252682979960075\\
0.135501084907905	0.00260509613089793\\
0.162601301889487	0.00278501889201236\\
0.189701518871068	0.00380836247137061\\
0.216801735852649	0.00691298330430388\\
0.24390195283423	0.0161171713628563\\
0.271002169815811	0.0406161637322368\\
0.298102386797392	0.105560111545878\\
0.325202603778973	0.262540056715896\\
0.352302820760554	0.603028995830709\\
0.379403037742135	1.24484577044734\\
0.406503254723716	2.2631100404726\\
0.433603471705298	3.57246314821606\\
0.460703688686879	4.84865161327449\\
0.48780390566846	5.62264187735486\\
0.514904122650041	5.55834415223577\\
0.542004339631622	4.68537240368484\\
0.569104556613203	3.38165876640722\\
0.596204773594784	2.10379486236148\\
0.623304990576365	1.14129478796206\\
0.650405207557946	0.547555864944028\\
0.677505424539527	0.237569321721899\\
0.704605641521108	0.0949994807826744\\
0.73170585850269	0.036459935376926\\
0.758806075484271	0.013806659806483\\
0.785906292465852	0.00557908066307613\\
0.813006509447433	0.00251455022881187\\
0.840106726429014	0.00143016182604477\\
0.867206943410595	0.000669427459769756\\
0.894307160392176	0.000846616565061644\\
0.921407377373757	0.00048121627354105\\
0.948507594355338	0.000899442505057199\\
0.975607811336919	0.000552501858758618\\
1.0027080283185	0.000446159725814096\\
1.02980824530008	0.000926281685499465\\
1.05690846228166	0.000442619156461387\\
1.08400867926324	0.000646520225048363\\
1.11110889624482	0.000596853258859279\\
1.13820911322641	0.000524665598065207\\
1.16530933020799	0.000717069577617184\\
1.19240954718957	0.00070807980906125\\
1.21950976417115	0.000639495370909211\\
1.24660998115273	0.000657344667026587\\
1.27371019813431	0.000637962265444782\\
1.30081041511589	0.000689649546659427\\
1.32791063209747	0.000748688402551427\\
};
\addlegendentry{$N=200; x=100; N/\sqrt{x}=20$}

\end{axis}
\end{tikzpicture}%
		\subcaption{PDFs for different lattice parameters}
		\label{fig:PDF_lattices}
	\end{minipage}
	\hfill
	\begin{minipage}[t]{0.49\textwidth}
		\centering
%
%
\begin{tikzpicture}

\begin{axis}[%
width=\figWidth,
height=\figHeight,
at={(0\figWidth,0\figHeight)},
scale only axis,
xmin=-1,
xmax=1,
xlabel style={font=\color{white!15!black}},
xlabel={\bjorx},
ymin=-10,
ymax=10,
ytick={-10,-5,...,10},
ylabel style={font=\color{white!15!black}},
ylabel={\pdf},
axis background/.style={fill=white},
xmajorgrids,
ymajorgrids,
legend style={at={(0.5,-0.16)}, anchor=north, legend cell align=left, align=left, draw=white!15!black},
ylabel style={{rotate=-90}}, scaled ticks=false, tick label style={/pgf/number format/fixed},
xticklabel style={yshift=-.5ex},
yticklabel style={xshift=-.5ex}
]
\addplot [color=black, dashed, line width=1.0pt, mark size=3.0pt, mark=+, mark options={solid, black}]
  table[row sep=crcr]{%
-2.61189944595093	0.000304610824835727\\
-2.50742346811289	0.000624034895658979\\
-2.40294749027485	0.00017937814845086\\
-2.29847151243682	0.000809884380558954\\
-2.19399553459878	2.20899964976773e-05\\
-2.08951955676074	0.00106066683563274\\
-1.98504357892271	-0.00020226167327858\\
-1.88056760108467	0.00134911322590727\\
-1.77609162324663	-0.000392378916771628\\
-1.67161564540859	0.00163907684264466\\
-1.56713966757056	-0.000580422366343993\\
-1.46266368973252	0.00197055116009434\\
-1.35818771189448	-0.000714622262219272\\
-1.25371173405645	0.00237662407458217\\
-1.14923575621841	-0.000852169824175792\\
-1.04475977838037	0.00272723118478137\\
-0.940283800542334	-0.00223641028405022\\
-0.835807822704297	-0.0351072446225506\\
-0.73133184486626	-0.410621447106723\\
-0.626855867028223	-1.79627532915628\\
-0.522379889190186	-3.36447316303583\\
-0.417903911352148	-2.64449956450312\\
-0.313427933514111	-0.829992785877725\\
-0.208951955676074	-0.0981260487513971\\
-0.104475977838037	0.00076503139961946\\
0	0.00347600999702598\\
0.104475977838037	0.00843135720139795\\
0.208951955676074	0.0975758856690269\\
0.313427933514111	0.806170379914477\\
0.417903911352148	2.59985175674942\\
0.522379889190186	3.36337578038173\\
0.626855867028223	1.82782970459931\\
0.73133184486626	0.426980826717625\\
0.835807822704297	0.0411076489420372\\
0.940283800542334	0.0042842679780783\\
1.04475977838037	-0.000316647961786124\\
1.14923575621841	0.00196240974712106\\
1.25371173405645	-0.000385254602371268\\
1.35818771189448	0.00165391995994872\\
1.46266368973252	-0.000319767609930771\\
1.56713966757056	0.00132838814189849\\
1.67161564540859	-0.000144312438768708\\
1.77609162324663	0.00108026342423517\\
1.88056760108467	-3.33494488293625e-05\\
1.98504357892271	0.000873862123336459\\
2.08951955676074	0.000147897979286355\\
2.19399553459878	0.000666385253765005\\
2.29847151243682	0.000339501112295349\\
2.40294749027485	0.000460881363800342\\
2.50742346811289	0.000530763638774618\\
};
\addlegendentry{m/g=2.8209}

\addplot [color=mycolor1, dashed, line width=1.0pt, mark size=3.0pt, mark=x, mark options={solid, mycolor1}]
  table[row sep=crcr]{%
-1.35466027924186	0.000838345290258488\\
-1.30047386807218	0.00054917968285069\\
-1.24628745690251	0.000889183571367978\\
-1.19210104573284	0.000583849356016363\\
-1.13791463456316	0.00101890678197238\\
-1.08372822339349	0.000605186035397446\\
-1.02954181222381	0.00113904578321627\\
-0.975355401054139	0.000644940650800777\\
-0.921168989884464	0.00136910510592388\\
-0.86698257871479	0.000419242667425688\\
-0.812796167545115	-0.000480151772440666\\
-0.758609756375441	-0.015392298029013\\
-0.704423345205767	-0.112933090943429\\
-0.650236934036092	-0.617258716733466\\
-0.596050522866418	-2.17348933727618\\
-0.541864111696744	-4.53333622828815\\
-0.487677700527069	-5.32596068831443\\
-0.433491289357395	-3.46706894397698\\
-0.379304878187721	-1.27355142309889\\
-0.325118467018046	-0.282630794444762\\
-0.270932055848372	-0.0409437361059881\\
-0.216745644678697	-0.00219063446759925\\
-0.162559233509023	0.00329048878393862\\
-0.108372822339349	0.00364256006871212\\
-0.0541864111696744	0.00409369251242122\\
0	0.00303999624407561\\
0.0541864111696744	0.0030254935239069\\
0.108372822339349	0.0024604182232816\\
0.162559233509023	0.00295032582235042\\
0.216745644678697	0.00771406899194706\\
0.270932055848372	0.0458549527511707\\
0.325118467018046	0.285103620743221\\
0.379304878187721	1.27250790444457\\
0.433491289357395	3.46180561843028\\
0.487677700527069	5.31527425404735\\
0.541864111696744	4.52315278820756\\
0.596050522866418	2.1718657211947\\
0.650236934036092	0.620569937764478\\
0.704423345205767	0.11700402920832\\
0.758609756375441	0.0177430672793768\\
0.812796167545115	0.00294454416766559\\
0.86698257871479	0.000942554065015977\\
0.921168989884464	0.000705392653500077\\
0.975355401054139	0.00054809391825387\\
1.02954181222381	0.000683781392219238\\
1.08372822339349	0.000501861228832388\\
1.13791463456316	0.000746711792978284\\
1.19210104573284	0.000478431540695429\\
1.24628745690251	0.000807807096234054\\
1.30047386807218	0.000492715716145149\\
};
\addlegendentry{m/g=5.6419}

\addplot [color=mycolor2, dashed, line width=1.0pt, mark size=3.0pt, mark=o, mark options={solid, mycolor2}]
  table[row sep=crcr]{%
-0.688296705648137	0.000500672679634913\\
-0.660764837422212	-0.0367157753486724\\
-0.633232969196286	-0.163487361172582\\
-0.605701100970361	-0.625008356103658\\
-0.578169232744435	-1.94492361500834\\
-0.55063736451851	-4.55071286020702\\
-0.523105496292584	-7.53667619664708\\
-0.495573628066659	-8.54196022792811\\
-0.468041759840733	-6.563240659059\\
-0.440509891614808	-3.47334208115211\\
-0.412978023388882	-1.32233352820743\\
-0.385446155162957	-0.388922022559439\\
-0.357914286937031	-0.0963299316084399\\
-0.330382418711106	-0.0208380389546361\\
-0.30285055048518	-0.00290177129282646\\
-0.275318682259255	0.00177775386613102\\
-0.247786814033329	0.00282523005065156\\
-0.220254945807404	0.0033829288570146\\
-0.192723077581478	0.00349768930268502\\
-0.165191209355553	0.0031487233598646\\
-0.137659341129627	0.00352255259650005\\
-0.110127472903702	0.00298782074257243\\
-0.0825956046777765	0.00297693600871759\\
-0.055063736451851	0.00289578212925763\\
-0.0275318682259255	0.00267586707145713\\
0	0.00260803746797012\\
0.0275318682259255	0.00249050315461925\\
0.055063736451851	0.0024828932782403\\
0.0825956046777765	0.00231755336262918\\
0.110127472903702	0.00242323377751496\\
0.137659341129627	0.00226063498377098\\
0.165191209355553	0.00242060391828463\\
0.192723077581478	0.00233040699537228\\
0.220254945807404	0.00263052731555485\\
0.247786814033329	0.00285193827172781\\
0.275318682259255	0.00423048228545008\\
0.30285055048518	0.00858964324843212\\
0.330382418711106	0.0271477754280534\\
0.357914286937031	0.105147510573769\\
0.385446155162957	0.409483468914089\\
0.412978023388882	1.37568682964523\\
0.440509891614808	3.57445635152657\\
0.468041759840733	6.66766385006467\\
0.495573628066659	8.55404782301635\\
0.523105496292584	7.44057140178282\\
0.55063736451851	4.43558927825869\\
0.578169232744435	1.87867983751951\\
0.605701100970361	0.601511361223313\\
0.633232969196286	0.158862792012941\\
0.660764837422212	0.0363357513060728\\
};
\addlegendentry{m/g=11.2838}

\end{axis}
\end{tikzpicture}%
		\subcaption{PDFs for different fermion masses}
		\label{fig:PDF_masses}
	\end{minipage}
	\caption{PDFs calculated with MPS. (\subref{fig:PDF_lattices}): \pdf for a mass of $\frac{m}{g}=5.6419 \approx \frac{10}{\sqrt{\pi}}$. (\subref{fig:PDF_masses}): \pdf for different masses of $\frac{m}{g} \in \{2.8209, 5.6419, 11.2838\} \approx\{5, 10, 20\} \big/\sqrt{\pi}$; parameters: $N=100;\, x=100;\, N/\sqrt{x}=10$.}
	\label{fig:PDF}
\end{figure}
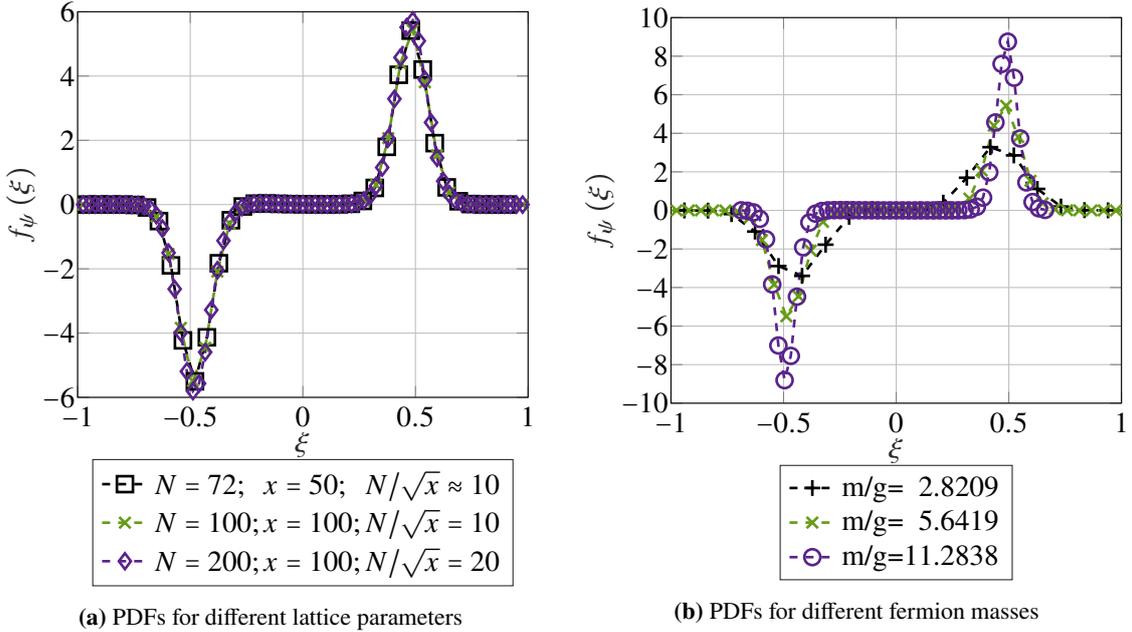

Finally, we obtain the PDF from the subtracted matrix elements by rescaling from $k$ to $\bjorx$ by a factor of $\frac{x}{m}$, where $m = E_h - E_0$ is the energy difference between the excited state and the ground state. The function \pdf obtained from the matrix elements in \cref{fig:CorrelatorSubtracted} is shown in \cref{fig:PDF_lattices}. A peak around $\bjorx \approx 0.5$ is present and is approximately symmetric with respect to the central value. The different lattice spacings and physical volumes lead to very similar results, which indicates that we are close enough to the continuum and thermodynamic limit. The systematic errors will be studied in more details in a future publication.

While the PDF is only defined for $\bjorx \in [0, 1]$, we can calculate \pdf outside this support as well. We find that \pdf vanishes for $|\bjorx| > 1$ as expected. Furthermore, \pdf is antisymmetric with respect to the axis $\bjorx = 0$, which is a consequence of the vanishing real part of the subtracted matrix elements (see \cref{fig:CorrelatorSubtracted_real}). This leads to a negative peak at around $\bjorx = -0.5$. Since \pdf for negative $\bjorx$ is related to the PDF of the antifermion by the relation $f_{\psi}(-\bjorx) = - \antipdf$~\cite{collins_2011}, the anti-symmetry means that $\pdf \equiv \antipdf$, i.e.\ the PDFs for the fermion and anti-fermion are identical.

These numerical results confirm the physical picture of the vector boson to be a meson, with symmetric roles of the fermion and the anti-fermion, i.e.\ both of these particles carry one half of the total meson momentum on average~\cite{SchwingerPartonPerturbative}. We note that the distributions are rather narrow in $\bjorx$-space, because of the large fermion mass.

We further calculate \pdf for different fermion masses. Examples are shown in \cref{fig:PDF_masses} for half and double the mass presented in \cref{fig:PDF_lattices}. We observe that the overall picture remains the same, but the peak widens (narrows) for lower (higher) mass. This is in good agreement with~\cite{SchwingerPDF_fastMovingFrame,SchwingerPDF_fastMovingFrameProceedings} and is expected as the distribution function becomes wider with decreasing mass. The limit of vanishing mass is known to be $\pdf \equiv 1$ for $\bjorx \in [0, 1]$, so the peak is widened maximally~\cite{SchwingerPartonPerturbative}. In the opposite limit, $m \rightarrow \infty$, the PDF should become a delta-function at $\bjorx = 0.5$, so maximally narrow.

\section{Conclusion and Outlook}
We demonstrated that it is possible to calculate PDFs directly from light-cone correlators with tensor networks. 
On the lattice, the light-cone direction can be replaced by a zigzag-like evolution of the Wilson line in space and time. The time evolution is possible with a Suzuki-Trotter decomposition and standard MPS-techniques like TEBD. For the spatial evolution, we introduced static charges that can be moved in order to evolve the electric flux step-wise. The PDF can be obtained by a Fourier transform after subtracting the ground state matrix element. While a detailed study of systematic errors will be published elsewhere, the results show good convergence with the lattice spacing and physical volume, indicating the feasibility of a continuum limit extrapolation.

Even though TNS methods can be applied in higher dimensions as well, the numerical costs increase. Moreover, the applicability of TNS methods to dynamical problems is limited by the entanglement growth. But the same strategy used for TNS calculations in the Hamiltonian formalism could in principle be implemented in quantum computers or simulators, which could potentially overcome these limitations. Indeed, first attempts have already been made to calculate Wilson lines in space and time in the context of quantum computing and with exact diagonalization for small systems~\cite{QC_PartonPhysics,QuantumSimulation,QS_lightfront,QC_PartonicCollinearStructure,QC_QuasiPDFs}.

In this work, we presented, for the first time, a full calculation of PDFs in the Schwinger model from light-cone correlators. We observed the correct symmetry properties and conclude that the fermion- and antifermion-PDF are identical for the vector boson, as expected for a meson. We further calculated the PDF for different fermion masses and found the expected broadening of the PDF with decreasing mass. These results demonstrate the applicability of TNS methods for more general LGT problems, beyond their previously established success for equilibrium states.

\acknowledgments
We would like to thank Hsiang-nan Li, Enrique Rico for fruitful discussions. 
This work was partly supported by  the DFG (German Research Foundation) under Germany's Excellence Strategy -- EXC-2111 -- 390814868, and Research Unit FOR 5522 (grant nr. 499180199), 
and by the EU-QUANTERA project TNiSQ (BA 6059/1-1), as well as Taiwanese NSTC grants 113-2119-M-007-013, 112-2811-M-A49-543-MY2, 112-2112-M-A49 -021-MY3 and 113-2123-M-A49-001.

\bibliography{bibliography}

\end{document}